\documentstyle[12pt]{article}

\def\hybrid{\topmargin 0pt	\oddsidemargin 0pt
	\headheight 0pt	\headsep 0pt
	\textheight 9in		
	\textwidth 6.25in	
	\marginparwidth .875in
	\parskip 5pt plus 1pt	\jot = 1.5ex}

\hybrid

\catcode`\@=11

\def\marginnote#1{}
\newcount\hour
\newcount\minute
\newtoks\amorpm
\hour=\time\divide\hour by60
\minute=\time{\multiply\hour by60 \global\advance\minute by-\hour}
\edef\standardtime{{\ifnum\hour<12 \global\amorpm={am}%
	\else\global\amorpm={pm}\advance\hour by-12 \fi
	\ifnum\hour=0 \hour=12 \fi
	\number\hour:\ifnum\minute<10 0\fi\number\minute\the\amorpm}}
\edef\militarytime{\number\hour:\ifnum\minute<10 0\fi\number\minute}

\def\draftlabel#1{{\@bsphack\if@filesw {\let\thepage\relax
   \xdef\@gtempa{\write\@auxout{\string
      \newlabel{#1}{{\@currentlabel}{\thepage}}}}}\@gtempa
   \if@nobreak \ifvmode\nobreak\fi\fi\fi\@esphack}
	\gdef\@eqnlabel{#1}}
\def\@eqnlabel{}
\def\@vacuum{}
\def\draftmarginnote#1{\marginpar{\raggedright\scriptsize\tt#1}}

\def\draft{\oddsidemargin -.5truein
	\def\@oddfoot{\sl preliminary draft \hfil
	\rm\thepage\hfil\sl\today\quad\militarytime}
	\let\@evenfoot\@oddfoot	\overfullrule 3pt
	\let\label=\draftlabel
	\let\marginnote=\draftmarginnote
   \def\@eqnnum{(\theequation)\rlap{\kern\marginparsep\tt\@eqnlabel}%
\global\let\@eqnlabel\@vacuum}  }

\def\preprint{\twocolumn\sloppy\flushbottom\parindent 2em
	\leftmargini 2em\leftmarginv .5em\leftmarginvi .5em
	\oddsidemargin -.5in	\evensidemargin -.5in
	\columnsep .4in	\footheight 0pt
	\textwidth 10in	\topmargin  -.4in
	\headheight 12pt \topskip .4in
	\textheight 7.1in \footskip 0pt
	\def\@oddhead{\thepage\hfil\addtocounter{page}{1}\thepage}
	\let\@evenhead\@oddhead	\def\@oddfoot{}	\def\@evenfoot{} }

\def\numberbysection{\@addtoreset{equation}{section}
	\def\theequation{\thesection.\arabic{equation}}}

\def\underline#1{\relax\ifmmode\@@underline#1\else
	$\@@underline{\hbox{#1}}$\relax\fi}

\def\titlepage{\@restonecolfalse\if@twocolumn\@restonecoltrue\onecolumn
     \else \newpage \fi \thispagestyle{empty}\c@page\z@
	\def\thefootnote{\fnsymbol{footnote}} }

\def\endtitlepage{\if@restonecol\twocolumn \else \newpage \fi
	\def\thefootnote{\arabic{footnote}}
	\setcounter{footnote}{0}}  

\catcode`@=12
\relax

\def\figcap{\section*{Figure Captions\markboth
	{FIGURECAPTIONS}{FIGURECAPTIONS}}\list
	{Figure \arabic{enumi}:\hfill}{\settowidth\labelwidth{Figure 999:}
	\leftmargin\labelwidth
	\advance\leftmargin\labelsep\usecounter{enumi}}}
 \relax
\def\tablecap{\section*{Table Captions\markboth
	{TABLECAPTIONS}{TABLECAPTIONS}}\list
	{Table \arabic{enumi}:\hfill}{\settowidth\labelwidth{Table 999:}
	\leftmargin\labelwidth
	\advance\leftmargin\labelsep\usecounter{enumi}}}
 \relax
\def\reflist{\section*{References\markboth
	{REFLIST}{REFLIST}}\list
	{[\arabic{enumi}]\hfill}{\settowidth\labelwidth{[999]}
	\leftmargin\labelwidth
	\advance\leftmargin\labelsep\usecounter{enumi}}}
 \relax
\makeatletter
\newcounter{pubctr}
\def\publist{\@ifnextchar[{\@publist}{\@@publist}}
\def\@publist[#1]{\list
	{[\arabic{pubctr}]\hfill}{\settowidth\labelwidth{[999]}
	\leftmargin\labelwidth
	\advance\leftmargin\labelsep
	\@nmbrlisttrue\def\@listctr{pubctr}
	\setcounter{pubctr}{#1}\addtocounter{pubctr}{-1}}}
\def\@@publist{\list
	{[\arabic{pubctr}]\hfill}{\settowidth\labelwidth{[999]}
	\leftmargin\labelwidth
	\advance\leftmargin\labelsep
	\@nmbrlisttrue\def\@listctr{pubctr}}}
 \relax
\makeatother
\newskip\humongous \humongous=0pt plus 1000pt minus 1000pt
\def\caja{\mathsurround=0pt}
\def\eqalign#1{\,\vcenter{\openup1\jot \caja
	\ialign{\strut \hfil$\displaystyle{##}$&$
	\displaystyle{{}##}$\hfil\crcr#1\crcr}}\,}
\newif\ifdtup
\def\panorama{\global\dtuptrue \openup1\jot \caja
	\everycr{\noalign{\ifdtup \global\dtupfalse
	\vskip-\lineskiplimit \vskip\normallineskiplimit
	\else \penalty\interdisplaylinepenalty \fi}}}
\def\eqalignno#1{\panorama \tabskip=\humongous
	\halign to\displaywidth{\hfil$\displaystyle{##}$
	\tabskip=0pt&$\displaystyle{{}##}$\hfil
	\tabskip=\humongous&\llap{$##$}\tabskip=0pt
	\crcr#1\crcr}}
\relax

\def\thefootnote{\fnsymbol{footnote}}
\def\ref#1{$^{#1)}$}
\def\a{\alpha}
\def\b{\beta}
\def\gg{\gamma}
\def\sl{$SL(2,R)_{k}/U(1)$ }
\def\su{$SU(2)_{N}/U(1)$ }
\def\c{\chi(z)}
\def\w{$W_{\infty}$ }
\def\ww{$W_{1+\infty}$ }
\def\wk{${\hat W}_{\infty}(k)$ }
\def\u{\underline}
\def\p{\psi}
\def\d{\Delta}
\def\pa{\partial}
\def\g{\Gamma}
\def\l{l_{1}}
\def\e{\epsilon}
\def\k{\sqrt{2\over k}}
\def\kk{\sqrt{k-2\over k}}
\def\cc{{\hat c}\,}
\def\ll{l_{2}}
\def\pt{\phi_{1}}
\def\pp{\phi_{2}}
\def\i{\infty}
\def\kl{\sqrt{2(k-2)}}
\begin{document}
\begin{titlepage}
\begin{center}
September 1991 \hfill    UCB-PTH-91/44 \\
            \hfill    LBL-31213 \\
            \hfill    UMD-PP-92-37\\
\vskip .1in

{\large \bf Beyond the Large N Limit: Non-linear $W_{\infty}$ as Symmetry}
\vskip .06in
{\large \bf of the SL(2,R)/U(1) Coset Model}
\footnote{This work was supported in part by the Director,
Office of
Energy Research, Office of High Energy and Nuclear Physics, Division of
High Energy Physics of the U.S. Department of Energy under Contract
DE-AC03-76SF00098 and in part by the National Science Foundation under
grants PHY-85-15857 and PHY-87-17155.}

\vskip .1in
\u{Ioannis Bakas}\footnote{Address after September 16, 1991:
School of Natural Sciences, Institute for Advanced Study, Princeton, NJ 08540,
USA.}\\
{\em Center for Theoretical Physics\footnote{e-mail:
BAKAS@UMDHEP.bitnet, BAKAS@UMDHEP.UMD.edu, UMDHEP::BAKAS}\\
Department of Physics and Astronomy\\
University of Maryland\\
College Park, MD 20742, USA}\\
\vskip .1in
and
\vskip .1in

\u{Elias Kiritsis}\footnote{ Address after October 1, 1991:
Lab. de Physique Th\'eorique, Ecole Normale Superieure, 24 rue Lhomond,
F-75231, Paris, CEDEX
05, FRANCE}\\

{\em  Department of Physics\footnote{e-mail:
THEORM::KIRITSIS, KIRITSIS@THEORM.LBL.Gov, KIRITSIS@LBL.bitnet},
      University of California and\\
      Theoretical Physics Group\\
      Lawrence Berkeley Laboratory\\
      Berkeley, CA 94720, USA}
\end{center}


\begin{abstract}
We show that the symmetry algebra of the $SL(2,R)_{k}/U(1)$ coset
model is a non-linear deformation of $W_{\infty}$, characterized by $k$.
This is a universal $W$-algebra which linearizes in the large $k$ limit and
truncates to $W_{N}$ for $k=-N$.
Using the theory of non-compact parafermions we construct a free field
realization of the non-linear $W_{\infty}$ in terms of two bosons with
background charge.
The $W$-characters of all unitary $SL(2,R)/U(1)$ representations
are computed.
Applications to the physics of 2-d black hole backgrounds are also discussed
and connections with the KP approach to $c=1$ string theory are outlined.
\end{abstract}
\end{titlepage}
\newpage
\renewcommand{\thepage}{\arabic{page}}
\setcounter{page}{1}

{\large\bf 1.} {\large\bf Introduction}
\bigskip

There has been considerable interest in the construction of a universal
$W$-algebra which unifies all types of extended conformal symmetries in 2-d
quantum field theory.
The existence of such master symmetry could be advantageous for developing a
non-perturbative formulation of string theory and exploring its vacuum
structure.
Originally it was thought that the large $N$ limit of Zamolodchikov's $W_{N}$
algebras, \cite{1,2}, interesting as it may be in its own right, plays a
prominent role in this direction.
However, since $W_{\infty}$ is a linear algebra, \cite{3,4,5}, it is very
difficult to invent a mechanism which effectively truncates the spin content of
$W_{\infty}$ to a finite set $2,3,\cdots ,N$ and produces the non-linear
features of $W_{N}$ algebras for $N \geq 3$.
To put it differently, since $W_{N}$ is not a subalgebra of $W_{N'}$ for $N<N'$
(unless $N=2$), the large $N$ limit is not defined inductively.

In this paper we show that the resolution to this problem lies beyond the large
$N$ limit!
The method we employ here is based on the standard relation between
$W$-algebras and parafermions.
Recall that the $Z_{N}$ parafermion coset models $SU(2)_{N}/U(1)$ have a
$W_{N}$
symmetry and central charge, \cite{6},
$$c_{N}={3N\over N+2}-1=2\,{N-1\over N+2}.\eqno(1.1)$$
For large values of $N$, $c_{N}$ approaches 2 from below and the underlying
$W_{N}$ algebra linearizes in the limit $N\rightarrow \infty$, \cite{5}.
One way to penetrate the $c=2$ barrier and go beyond the large $N$ limit of
$SU(2)$
parafermionic models is to consider the non-compact coset models $SL(2,R)
_{k}/U(1)$, \cite{7,8}, whose central charge is
$$c_{k}={3k\over k-2}-1=2\,{k+1\over k-2}.\eqno(1.2)$$
Then, it is natural to expect that the $W$-algebra of the \sl coset models
reduces to the $W_{N}$ algebra of \su for $k=-N$.

The idea to make the transition from the compact to the non-compact models
is motivated by well known results in the representation theory of the $N=2$
superconformal algebra.
In this case, for central charge $c<3$, there exists only a discrete series of
unitary representations with $c=3N/(N+2)$, for all non-negative integers $N$,
\cite{9}.
These can be obtained from representations of the $SU(2)_{N}$ current algebra
by subtracting and then adding back a free boson (alternatively stated by
changing the radius of the torus boson), \cite{10}.
For representations with $c>3$, the SU(2) parafermion method is not adequate
and further generalization is required.
Lykken introduced a new type of parafermion algebra for $c>3$, \cite{11},
which corresponds to $SL(2,R)$ and contains an infinite number
of parafermions, unlike the compact case.
It was subsequently shown that all unitary $N=2$ superconformal representations
with $3<c=3k/(k-2)$ can be obtained from representations of the
$SL(2,R)_{k}$
current algebra by the same method of appropriately subtracting and then adding
back a free boson, \cite{7}.

Our main result is the construction of the $W$-symmetry of the \sl coset model.
As we will demonstrate later, this is an extended conformal algebra which for
generic real values of $k$ has an infinite number of generators with integer
spin $2,3,\cdots$.
Although the spectrum of generators is identical to that of the $W_{\infty}$
algebra,
in this case there are non-linear terms which depend explicitly on $k$ and
cannot be absorbed into field redefinitions.
We will denote the $W$-algebra of the \sl coset model by \wk and note that the
non-linearities disappear in the limit $k\rightarrow \pm\infty$, where the
ordinary \w algebra is recovered.
Also, for all integer values of $k=-N\leq -2$, it turns out that \wk truncates
to the finitely generated $W_{N}$ algebras.
In this sense, the non-linear deformation of \w we construct here is a
universal $W$-algebra for the $A$-series of extended conformal symmetries.
Similarly, one may consider higher rank non-compact coset spaces and obtain
further generalizations of our results.
We will comment on this possibility at the end.

The $W$-symmetry of the \sl model (and its parafermionic extensions that we
will also consider for rational values of $k$) are also
important for studying properties of the conformal field theory governing the
2-d black hole solution to string theory discovered by Witten, \cite{12} (see
also \cite{13}).
In this case $k=9/4$ $(c=26)$ and the \wk symmetry is non-linear.
However, since the number of its generators is infinite, the 2-d black hole
background carries an infinite number of independent quantum numbers, one for
each $W$-generator.
We will propose a recursive method for their construction.

Another problem related to the recent developments in non-perturbative
2-d quantum gravity, \cite{14} is the connection (if any) between \wk
(for $k=9/4$) and the
underlying $W_{1+\infty}$ symmetry in $c=1$ string theory, \cite{15}.
It has been demonstrated recently that the bi-Hamiltonian structure of
the full KP hierarchy can be described in $W$ terms as follows:
$W_{1+\infty}$, \cite{16} provides the first Hamiltonian structure,
\cite{17,18}, while the second one is a non-linear deformation of \w,
\cite{19,20} (in both cases the central charge is zero).
It is reasonable to expect that \wk is a quantum version of the second
Hamiltonian structure of the KP hierarchy.
{}From the point of view of the KP approach to the theory of multi-matrix
models,
\cite{21}, the relation between the 2-d black hole solution and the $c=1$
matrix model could be elucidated and lead to a deeper understanding of these
string models.
We will present some thoughts in this direction later.

The material in this paper is organized as follows.
In section 2 we review the basics of parafermion algebras with emphasis on the
$SL(2,R)$ case and motivate the introduction of \wk.
In section 3 we adopt the free field realization of $SL(2,R)_{k}$ current
algebra and obtain expressions for the generators of \wk in terms of two
bosons with background charge.
We also present the results of some explicit calculations for the commutation
relations of \wk, which are sufficient to demonstrate the universal features of
the algebra.
In section 4 we derive the character formulae for all unitary highest weight
representations.
In section 5 we discuss the implications of our results to the theory of 2-d
gravity and finally in section 6 we present our conclusions.

\bigskip
{\large\bf 2.} {\large\bf SL(2,R)/U(1) Parafermion Algebra}
\bigskip

Parafermion algebras are generated by a collection of currents $\p_{l}(z)$ and
their conjugates $\p_{l}^{\dag}(z)\equiv\p_{-l}(z)$ with
$\p_{0}(z)=\p_{0}^{\dag}(z)=1$.
A priori there are no restrictions on the range of $l=0,1,2,\cdots$ and
generically, the conformal dimension $\Delta_{l}$ of the parafermion currents
is fractional.
The algebra assumes the form, \cite{11},
$$\p_{\l}(z)\p_{l_{2}}(w)=C_{l_{1},l_{2}}(z-w)
^{\Delta_{l_{1}+l_{2}}-\Delta_{l_{1}}-\Delta_{l_{2}}}\left[ \p_{l_{1}+l_{2}}(w)
+{\cal O}(z-w)\right],\eqno(2.1a)$$

$$\p_{l}(z)\p_{l}^{\dag}(w)=(z-w)^{-2\Delta_{l}}\left[1+{2\Delta_{l}\over
c_{\p}}(z-w)^{2}T_{\p}(w)+{\cal O}((z-w)^3)\right],\eqno(2.1b)$$
where $T_{\p}$ is the stress tensor of the parafermion theory with central
charge $c_{\p}$.
The structure constants $C_{l_{1},l_{2}}$ are determined by associativity and
the conformal dimensions $\Delta_{l}$ are constrained to satisfy the recursion
relation
$$\Delta_{l+1}+\Delta_{l-1}-2\Delta_{l}-\Delta_{2}+2\Delta_{1}=n_{l}
\,\,\,,\,\,\,n_{l}\in Z^{+}_{0}\;.\eqno(2.2)$$
Parafermions arise naturally in the context of the $N=2$ superconformal
algebra,
\cite{10,11}.
Its generators, $J(z)$, $T(z)$ and $G^{\pm}(z)$ can be constructed in terms of
one scalar field $\varphi(z)$ and two (non-local) parafermion currents
as follows:
$$J(z)=i\sqrt{c\over 3}\pa_{z}\varphi(z)\,\,\,,\,\,\,T(z)=-{1\over 2}(\pa_{z}
\varphi(z))^{2}+T_{\p}(z)\,\,\,,\eqno(2.3a)$$
$$ G^{\pm}(z)=\sqrt{2c\over 3}e^{\pm i\sqrt{3\over c}\varphi(z)}\p_{\pm 1}(z).
\eqno(2.3b)$$
This realization determines $c_{\p}$ and $\Delta_{1}$ in terms of the central
charge of the $N=2$ superconformal algebra,
$$c_{\p}=c-1\,\,\,,\,\,\,\Delta_{1}={3(c-1)\over 2c}\,\,\,\eqno(2.4)$$
and the operator product expansion (OPE) of the $l=1$ parafermions generates
$\p_{\pm 2}$ with conformal dimension
$$\Delta_{2}={2(2c-3)\over c}\;.\eqno(2.5)$$

It is clear that the spectrum of the parafermion algebra depends crucially on
$c$.
In particular, for $c<3$, using the parametrization $c=3N/(N+2)$, the recursion
relations (2.2) (due to zeros in the OPE coefficients $C_{l_{1},l_{2}}$)
generate
only a finite number of independent parafermion currents with dimensions
$$\d_{l}={l(N-l)\over N}\,\,\,,\,\,\,0\leq l\leq N-1\;.\eqno(2.6)$$
These correspond to the ordinary $Z_{N}$ parafermions of Fateev and
Zamolodchikov, \cite{6} described by the coset model \su, with central charge
$c_{\p}$ given by eq. (1.1).
On the other hand, for $c\geq 3$, using the parametrization $c=3k/(k-2)$,
Lykken
pointed out, \cite{11} that the parafermion algebra does not truncate.
In this case the parafermion currents form an infinite family of fields with
dimensions
$$\d_{l}={l(k+l)\over k}\,\,\,,\,\,\,l\in Z^{+}_{0}\;.\eqno(2.7)$$
For completeness we include the formulae for the structure constants $C_{\l,
\ll}$ in both cases.
For the $Z_{N}$ parafermions,
$$C_{\l,\ll}=\left[{\g(N-\l+1)\g(N-\ll+1)\g(\l+\ll+1)\over \g(\l+1)
\g(\ll+1)\g(N+1)\g(N-\l-\ll+1)}\right]^{1\over 2}\,,\,\eqno(2.8)$$
while for the Lykken family,
$$C_{\l,\ll}=\left[{\g(k+\l+\ll)\g(k)\g(\l+\ll+1)\over \g(\l+1)\g(
\ll+1)\g(k+\l)\g(k+\ll)}\right]^{1\over 2}\,.\eqno(2.9)$$
We also have $C_{-\l,\ll}=C_{\l,\ll-\l}$ for $\ll\geq \l$.
Notice that eqs. (2.6), (2.8) are analytic continuations of eqs. (2.7), (2.9)
respectively for $k\rightarrow -N$.

The parafermion algebra (2.1) can be converted into a current algebra using an
additional free scalar field $\c$, \cite{6}, in analogy with the $N=2$
superconformal algebra.
To do this, we define the currents
$$J^{3}(z)=-\sqrt{k\over 2}\,\pa_{z}\c\,\,\,,\,\,\,J^{\pm}(z)
=\sqrt{k}\,e^{\pm\sqrt{2\over k}\c}\p_{\pm 1}(z)\;,\eqno(2.10)$$
which transform the parafermion algebra into the current algebra
$$J^{+}(z)J^{-}(w)={k\over (z-w)^{2}}-2\,{J^{3}(w)\over z-w}+{\cal O}(1)\,,
\eqno(2.11a)$$
$$J^{3}(z)J^{\pm}(w)=\pm{J^{\pm}(w)\over z-w}+{\cal O}(1)\,,\eqno(2.11b)$$
$$J^{3}(z)J^{3}(w)=-{k/2\over (z-w)^2}+{\cal O}(1)\,\,.\eqno(2.11c)$$
Here we consider the case of Lykken parafermions with $c_{\p}\geq 2$
(i.e., $k$ is any real number $\geq 2$).
The symmetry of the non-compact parafermions is $Z_{\i}$.
The OPEs (2.11) define an $SL(2,R)_{k}$ current algebra and the
corresponding coset model is \sl with central charge $c_{\p}$ given by eq.
(1.2).
If we had taken $k=-N$, we would have obtained the $SU(2)_{N}$ current algebra
description of the $Z_{N}$ parafermions, \cite{6}.
The free field representation
of $Z_{N}$ parafermions was derived in \cite{22}.
The corresponding one for the non-compact case was given in
\cite{23}.

It is well known that parafermion coset models have a $W$-symmetry whose
generators appear on the right hand side of the OPE (2.1b).
In particular, it is sufficient to consider $\p_{1}(z)\p_{-1}(w)$ and go beyond
the most singular terms in order to extract the $W$-generators.
In eq. (2.1b) we have only displayed explicitly the stress tensor $T_{\p}$
which
is the lowest spin field of the underlying $W$-algebra.
For $SU(2)_{N}$ parafermion models the corresponding $W$-algebra is $W_{N}$,
\cite{1,2}, generated by chiral fields with integer spin $2,3,\cdots,N$.
For finite $N\geq 3$, the parafermion currents acquire fractional dimensions
$\sim 1/N$ and the $W_{N}$ algebra is non-linear.
However, as $N\rightarrow\infty$, the parafermions become mutually local fields
with integer dimension $\d_{l}=1,2,\cdots $ and \w linearizes, \cite{5}.
Then, by formally interchanging $N$ to $-k$ in the large $N$ limit,
we access the non-compact coset region which lies beyond the $c=2$ barrier.
For $c\geq 2$, the \sl model has the same number of parafermion currents
as $SU(2)_{\infty}/U(1)$ and makes sense for any real $k>2$.

In order to understand qualitatively the main features of the \sl $W$-algebra,
notice that although the number of parafermions is independent of $k$, their
conformal dimension $\d_{l}=l+{l^2\over k}$ acquires a non-integer contribution
$\sim 1/k$ for finite $k$.
Therefore, it is natural to expect that the underlying $W$-algebra has the same
spectrum of generators as \w, but it develops non-linearities depending on $k$.
In the next section we construct the $W$-algebra of the \sl coset model, \wk
and demonstrate its universal features.
In section 4 we show that the character of the \wk vacuum module (as well
as the rest of the characters) for $k>2$ are independent of $k$.
This rigorously proves that the spectrum of the algebra is the same as
that of the usual \w.
Thus, for any $k>2$, the generators are chiral fields with integer spin $2,3,
\cdots,$ each with multiplicity one.
\newpage
{\large\bf 3.} {\large\bf The Chiral Algebra \wk}
\bigskip

We adopt the free field realization of the $SL(2,R)_{k}$ current algebra,
which represents the two parafermion currents $\p_{\pm 1}(z)$ in eq. (
2.10) as, \cite{22,23},
$$\p_{\pm 1}(z)={1\over \sqrt{2k}}\,\,\left[\mp\sqrt{k-2}\,\pa_{z}\pt(z)
+i\sqrt{k}\,\,\pa_{z}\pp(z)\right] e^{\pm i\sqrt{2\over k}\pp(z)}\;,
\eqno(3.1)$$
using two free bosons with
$$\langle \phi_{i}(z)\phi_{j}(w)\rangle=-\delta_{ij}\log(z-w)\,\,\,.\,\,\,
\eqno(3.2)$$
It follows immediately from the OPE (2.1b) that the stress tensor of the theory
is
$$W_{2}(z)\equiv T_{\p}(z)=-{1\over 2}(\pa_{z}\pt)^{2}-{1\over 2}
(\pa_{z}\pp)^{2}+{1\over \sqrt{2(k-2)}}\,\pa^{2}_{z}\pt\,\,.\eqno(3.3)$$
All other generators of \wk appear in the less singular terms of $\psi_{1}(z)
\p_{-1}(w)$ and it is just a matter of computation to extract their form in
terms of $\pt,\pp$.
We will present the results of the calculation up to spin 5, since for
the higher spin generators the expressions become considerably more involved
and we have no closed formulae for them in general.
It should be stressed, however, that the OPEs of the $W$-algebra can be
computed without the use of the free-field formulation, directly from
the parafermion correlation functions. In several occasions we did
the computations with both methods as an independent check.

We define the $primary$ higher spin generators $W_{s}(z)$ of the algebra,
using the bootstrap method, by the OPE\footnote{Normal ordered products
of operators are defined, as usual, by subtracting the singular terms plus
the finite terms that are total derivatives of lower dimension operators
appearing in the corresponding OPE. In general, they are not symmetric.}
$$\eqalign{&\p_{1}(z+\e)\p_{-1}(z)=\e^{-2{k+1\over k}}\Biggl[1+{k-2\over k}
\biggl(\e^{2}+{1\over 2}\e^{3}\pa_{z} +{3\over 20}\e^{4}\pa_{z}^{2}+
{1\over 30}\e^5\pa_{z}^{3}\biggr)W_{2}(z)-\cr
&-{1\over 4}\biggl(\e^3 +{1\over 2}\e^4\pa_{z} +{1\over 7}\e^5
\pa_{z}^{2}\biggr) W_{3}(z)+{(6k+5)(k-2)^2\over 2k^{2}(16k-17)}
\left(\e^4 +{1\over 2}\e^5\pa_{z}\right):W_{2}^{2}:(z)+\cr
&+ {1\over 32}\left(\e^{4}+{1\over 2}\e^5\pa_{z} \right)W_{4}(z)
-{1\over 3\cdot2^{7}}\e^{5}W_{5}(z)-{(10k+7)(k-2)\over 4k(64k-107)}\e^{5}
:W_{2}W_{3}:(z)+{\cal O}(\e^{6})\Biggr].\cr}\eqno(3.4)$$

For the primary spin 3 field we obtain the expression
$$W_{3}(z)=\alpha (\pa\pp)^{3}+\beta\pa^{3}\pp+\gamma(\pa\pt)^{2}\pa\pp+\delta
\pa^{2}\pt\pa\pp+\varepsilon\pa\pt\pa^{2}\pp\,,\eqno(3.5)$$
where
$$\alpha={2i\over 3}{(3k-4)\over k}\sqrt{2\over k}\,\,\,,\,\,\,\beta={i\over 3}
\k\,\,\,,\,\,\,\gg =2i{(k-2)\over k}\k\;,$$
$$\delta=2i{(k-2)\over k}\kk\,\,\,,\,\,\,\varepsilon=-2i\kk\;.\eqno(3.6)$$
Then, the OPE of $W_{3}$ with itself yields
$$\eqalignno{W_{3}(z+\e)W_{3}(z)&={16\over 3}{(k+1)(k+2)(3k-4)\over
k^{3}\e^{6}}+
{16(k+2)(k-2)(3k-4)\over k^{3}}\biggl({1\over \e^{4}}+\cr
&+{1\over 2}{\pa_{z}\over
\e^{3}}+{3\over 20}{\pa^{2}_{z}\over \e^2}+{1\over 30}{\pa^{3}_{z}\over \e}
\biggr)W_{2}(z)+{2(2k-3)\over k}\biggl({1\over \e^2}+{1\over 2}{\pa_z \over\e}
\biggr)W_{4}(z)+\cr
&+{2^{7}(k+2)(3k-4)(k-2)^{2}\over
k^{3}(16k-17)}\biggl({1\over \e^{2}}+{1\over 2}{\pa_{z}\over \e}\biggr)
:W_{2}^{2}:(z)+{\cal O}(1)\;,&(3.7)\cr}$$
where $W_{4}(z)$ is the primary spin 4 field which was defined implicitly in
eq. (3.4).
Its explicit form is given in appendix A.

Before we proceed further a remark is in order.
The requirement that a spin 3 field of the form (3.5) is primary with
respect to the stress tensor (3.3), imposes three conditions on its numerical
coefficients.
Since the overall normalization is a matter of convention, one is left with one
undetermined parameter.
Then, if we consider the OPE $W_{3}(z+\e)W_{3}(z)$ and demand that the
$\e^{-4}$
term is proportional to $W_{2}(z)$, we obtain a quadratic equation for the
ratio $x\equiv \gamma/\varepsilon$.
The two solutions are
$$x_{+}=-{\sqrt{2(k-2)}\over k}\,\,\,,\,\,\,x_{-}=-{2\over 3}\sqrt{2(k-2)
}\,\,.\,\,\eqno(3.8)$$
In our case the expression (3.6) corresponds to the first branch, which also
gives rise to a non-zero primary field $W_{4}$.
For the second branch, the only quasiprimary operator that appears to order
$\e^{-2}$ is $:W_{2}^{2}:$ and there is no primary spin 4 field present.
In this case we are dealing with Zamolodchikov's $W_{3}$ algebra in its free
field realization in terms of two bosons, \cite{1}.

We return now to the \wk algebra and perform the next OPE. We find
$$\eqalignno{W_{3}(z+\e)W_{4}(z)&={3\cdot 2^{7}(k+3)(2k-1)(2k-3)\over k^{2}
(16k-17)}\biggl({1\over \e^{4}}+{1\over 3}{\pa_{z}\over \e^{3}}+
{1\over 14}{\pa^{2}_{z}\over \e^{2}}+\cr
&+{1\over 84}{\pa^{3}_{z}\over \e}\biggr)W_{3}(z)+{5(3k-4)\over 3k}\biggl(
{1\over \e^{2}}+{2\over 5}{\pa_{z}\over \e}\biggr)W_{5}(z)+\cr
&+{39\cdot 2^{8}(k-2)(k+3)(2k-1)(2k-3)\over k^{2}(16k-17)(64k-107)}\biggl(
{1\over \e^{2}}+{2\over 5}{\pa_{z}\over \e}\biggr):W_{2}W_{3}:(z)-\cr
&-{3\cdot 2^{6}(k+3)(2k-3)(k-2)\over k^{2}(16k-17)}{:(\pa W_{2})W_{3}:(z)\over
\e}+{\cal O}(1)\;,&(3.9)\cr}$$
where $W_{5}(z)$ is the spin 5 primary field computed directly from the
expansion (3.4).
Its explicit form is also given in appendix A.
The expressions for $:W_{2}^{2}:$,
$:W_{2}W_{3}:$ and $:(\pa W_{2})W_{3}:$ are given in appendix B.

We may iterate the procedure above to obtain expressions for all higher spin
fields of the algebra and compute their commutation relations.
The strategy is obvious, but the derivation of the formulae for all the higher
spin fields in closed form is a very difficult task.
As far as the commutation relations of \wk are concerned, one may compute its
central terms (i.e., $\langle W_{s}(z)W_{s'}(w)\rangle$) directly from eq.
(3.4)
by considering the parafermion four-point function
$$\langle \p_{1}(z_{1})\p_{1}^{\dag}(z_{2})\p_{1}(z_{3})\p_{1}^{\dag}(
z_{4})\rangle =\left[{z_{13}z_{24}\over
z_{12}z_{14}z_{34}z_{23}}\right]^{2\over
k}\Biggl[{1\over z_{12}^{2}z_{34}^{2}}\biggl(1-{2\over k}{z_{34}z_{12}\over
z_{24}z_{23}}\biggr)+(z_{2}\leftrightarrow z_{4})\Biggr].\eqno(3.10)$$
The coefficients of all other terms that appear in the singular part of the OPE
$W_{s}(z+\e)$$W_{s'}(z)$ can also be computed directly, using eq. (3.4) and the
parafermion six-point function $\langle \p_{1}\p_{1}^{\dag}\p_{1}\p_{1}^{\dag}
\p_{1}\p_{1}^{\dag}\rangle$.

A general property of \wk is already obvious from (3.7), (3.9).
The algebra is invariant under the transformation $W_{s}\rightarrow (-1)
^{s}W_{s}$, where composite fields transform multiplicatively.
This is a reflection of the $\p_{l}\rightarrow \p_{-l}$ invariance of the
parafermion algebra and corresponds to the Weyl reflection in $SL(2,R)$.
We note that the closure of the \wk algebra is guaranteed by the closure of
the charge zero sector of the enveloping algebra of the $SL(2,R)_{k}$
current
algebra (c.f. the realization (2.10) in terms of $\p_{1}$, $\p_{1}^{\dag}$
and the additional scalar field $\chi$).
It would be very interesting, though, to have the complete structure of \wk
in closed form.

For all real values of $k> 2$, the \wk algebra is non-linear and has an
infinite number of higher spin generators (one per spin).
One might wonder that the algebra, although non-linear at first sight, might
linearize by appropriate change of basis. This was the case with the standard
\w.
For example, one might consider the field redefinition
$${\tilde W}_{4}(z)=W_{4}(z)+{64(k+2)(3k-4)(k-2)^{2}\over k^{2}(2k-3)(
16k-17)}:W_{2}^{2}:(z)\,\,,\,\,\eqno(3.11)$$
which linearizes the singular part of the OPE (3.7) and still maintains the
linearity of the $W_{2}{\tilde W}_{4}$ OPE.
In order to verify that such field redefinition does not remove all the
non-linearities from \wk, it is sufficient to look at the OPE $W_{4}W_{4}$
and in particular to order $\e^{-4}$, where both $W_{4}$ and $:W_{2}^{2}:$
terms appear.
We find that
$$\eqalign{W_{4}(z&+\e)W_{4}(z)={2^{10}(k+1)(k+2)(k+3)(2k-1)(3k-4)\over k^{4}(
16k-17)\e^{8}}\cdot\cr
&.\Biggl[1+{4(k-2)\over (k+1)}\biggl(\e^{2}+{1\over
2}\e^{3}\pa_{z}+{3\over 20}\e^{4}\pa^{2}_{z}\biggr)W_{2}(z)
+{42(k-2)^{2}\over (k+1)(16k-17)}\e^{4}:W^{2}_{2}:(z)+\cr
&+{9k^{2}(4k^3+15k^2-33k+4)\over 16(k+1)(k+2)(k+3)(2k-1)(3k-4)}\e^{4}W_{4}(z)
+{\cal O}(\e^{5})\Biggr].\cr}\eqno(3.12)$$
Then, if we compute ${\tilde W}_{4}{\tilde W}_{4}$, we find to order $\e^{-4}$
that the coefficients of $W_{4}$ and $:W_{2}^{2}:$ do not combine to give only
${\tilde W}_{4}$. It turns out that the remaining term is
$${160(k-1)(k-2)^{2}(k+2)(3k-4)\over 3k^{2}(2k-3)(18k^{3}+k^{2}-117k+116)}:
W_{2}^{2}:(z)\;,\eqno(3.13)$$
which establishes the non-linear nature of \wk.

Next, we discuss properties of \wk for some special values of $k$.

(i) \u{{\em Recovering} \w}:  In the limit $k\rightarrow \infty$, the
$W$-symmetry of the
\sl coset model is the standard \w, since as we argued earlier the compact
and non-compact parafermion algebras have a common limit.
In both cases the algebra contracts to a $U(1)^{3}$ current
algebra\footnote{In the $SL(2,R)$ case two of the abelian currents
are non-compact.}
by a suitable rescaling of its generators. Passing to the coset model, one
effectively removes one of the $U(1)$ currents and the resulting parafermion
algebra is the enveloping algebra of the $U(1)^{2}$ current algebra, \cite{5}.
The first parafermion currents can be identified with the two $U(1)$ currents,
which when written in terms of a free complex boson are
$$\p_{1}(z)=i\pa \phi(z)\,\,\,,\,\,\,\p_{1}^{\dag}(z)=i\pa{\bar \phi}(z)\,\,.
\eqno(3.14)$$
All higher parafermions are composite operators,
$$\p_{k}(z)=:\p_{1}^{k}:(z)\,\,\,,\,\,\,\p_{k}^{\dag}(z)=:(\p_{1}^{\dag})^{k}
:(z)\,\,.\eqno(3.15)$$

In the large $k$ limit, the free field realization of \wk in terms of
two bosons with background charge should reproduce the results we have obtained
for \w with $c=2$, \cite{5}.
This can be readily verified using the identification
$$\pt(z)={1\over \sqrt{2}}(\pa\phi(z)+\pa{\bar \phi}(z))\,\,\,,\,\,\,\pp(z)
={i\over \sqrt{2}}(\pa \phi(z)-\pa{\bar \phi}(z))\,\,,\,\,\eqno(3.16)$$
which yields
$${\tilde W}_{2}(z)\equiv W_{2}(z)=-\pa\phi\pa{\bar \phi}(z)\eqno(3.17)$$
at $k=\infty$.
Similarly, $W_{3}(z)$ becomes
$${\tilde W}_{3}(z)\equiv W_{3}(z)=-2(\pa\phi\pa^{2}{\bar \phi}-
\pa^{2}\phi\pa{\bar \phi})(z)\,\,.
\eqno(3.18)$$
For the higher spin fields we proceed along similar lines.
We note, however, that \w is usually written in a quasiprimary basis, while all
our \wk generators are by construction primary.
In order to make the comparison for $s=4$, we should consider ${\tilde W}_{4}$
instead of $W_{4}$. Using eqs. (3.10), (A.1), (A.2) and (B.1) we obtain at
$k=\infty$
$${\tilde W}_{4}(z)=-{16\over 5}(\pa\phi\pa^{3}{\bar \phi}
-3\pa^{2}\phi\pa^{2}{\bar \phi}+\pa^{3}\phi\pa{\bar \phi})(z)\,\,.\,\,
\eqno(3.19)$$
These are exactly the formulae we have given elsewhere, \cite{5} for the \w
algebra and they generalize for higher spin too.
The general expression for the appropriate ${\tilde W}_{s}$ can be obtained
from the following expansion
$$\p_{1}(z+\e)\p_{-1}(z)=\e^{-2}\Biggl[1+\sum_{s=2}^{\infty}{(-1)
^{s}\e^{s}\over (s-2)!2^{2s-4}}\sum_{n=0}^{\infty}{(2s-1)!(s+n-1)!
\over n!(s-1)!(2s+n-1)!}\e^{n}\pa^{n}_{z}{\tilde W}_{s}(z)+{\cal O}
({1\over k})\Biggr].\eqno(3.20)$$
The linearization of the \wk algebra at $k=\infty$ can also be seen explicitly
in the example we gave earlier, since the difference (3.13) vanishes in this
limit.

(ii) \u{\wk $at$ $k=2$}:  So far we have been mostly interested in \wk for
real $k>2$.
There is no a priori reason not to extend the range of $k$ to all real numbers.
The value $k=2$ is special because the central charge of the Virasoro
subalgebra becomes infinite.
Nevertheless, it is interesting to see what happens to the $W$-symmetry there.
In the examples we have already given, all non-linear terms disappear and there
no need to make any field redefinitions.
Moreover, the stress tensor and its composites do not appear in the commutation
relations of higher spin generators which are well defined with no infinities
involved.
This can be easily understood by looking at the free field realization of
$W_{3}(z)$, $W_{4}(z)$, $W_{5}(z)$, etc.
For $k=2$, all terms involving the field $\pt(z)$ vanish and therefore the
remaining field $\pp(z)$ is insensitive to the infinite background charge.
In this case, because of the decoupling that occurs, we may disregard $W_{2}(z)
$ and consider the subalgebra of ${\hat W}_{\infty}(2)$ generated by all other
higher spin fields. The latter is not conformal, but it is consistent with the
Jacobi identities, which makes it worthwhile for further investigation.

We expect that the linearization of the algebra at $k=2$ persists
for all $s\geq 3$.
The reason is that for $k=2$, the dimension of the parafermion currents (2.7)
assumes only integer or half-integer values.
Experience suggests that when the parafermions are mutually
local, the corresponding $W$-algebra is linear.
We can also provide a general argument for the decoupling of any composite
operator that contains the stress tensor.
The norm of such operators is a polynomial in the central charge of the
Virasoro subalgebra with degree greater than zero.
These operators appear in the OPE of $W$-generators with coefficients
inversely proportional to their norm.
Thus, they will vanish when the central charge becomes infinite.
We note that the existence of linear higher spin algebras with no Virasoro
generators has been suggested in a different context before, \cite{17}, using a
twisted version of $W_{1+\infty}$\footnote{The quantum version of these
algebras is described in \cite{25}.}.
Whether ${\hat W}_{\infty}(2)$ minus the Virasoro subalgebra has any
relation with them remains to be seen.

(iii) \u{$Truncation$ $to$ $W_{N}$}:  Consider now \wk with $k<2$.
Since the construction of our algebra is based on parafermions, we know in
advance that for $k=-N=-2,-3,-4,\cdots$ the infinite structure should truncate
and the ordinary theory of $Z_{N}$ parafermions and $W_{N}$ algebras should be
recovered.
This is illustrated very nicely in our examples.
Consider first $k=-2$ and notice that apart from the stress tensor all other
$W$-generators are null, i.e., $\langle W_{3}(z)W_{3}(w)\rangle =\langle W_{4}
(z)W_{4}(w)\rangle=\cdots=0$.
Therefore, it is legitimate to factor them out and reduce ${\hat W}_{\infty}(
-2)$ to the Virasoro algebra as required.
Similarly, for $k=-3$, all $W$-generators are null apart from $W_{2}(z)$ and
$W_{3}(z)$ and upon reduction, Zamolodchikov's $W_{3}$ algebra emerges.
The norm of $W_{s}$ is proportional to $\prod_{i=1}^{s-1}(k+i)$ .
This has been explicitly checked up to $s=5$ (see appendix C).
All $W_{N}$ algebras can be obtained in this fashion, which makes \wk a
universal $W$-algebra.

To avoid confusion we emphasize that the reduction of ${\hat
W}_{\infty}(-N)$ to $W_{N}$ is not obtained because all other structure
constants of the algebra, apart from those of $W_{N}$, vanish.
It can be readily seen that this is not the case.
The precise statement is that all generators with $s>N$ are null and they
generate an ideal of the algebra, in the sense that the OPE of any operator
with a null operator will produce only null operators.
Thus, $W_{N}$ is of cohomological nature, defined as the original algebra
modulo the null ideal.
This point of view is also adopted by Narganes-Quijano, \cite{22},
while discussing the free-field realization of $W_{N}$ algebras
in terms of two bosons.

In the region $-1<k<2$, where the central charge is negative, we also
observe a reduction of the \wk algebra for some fractional values of $k$,
like $1/2,4/3,\cdots$ etc.
The identification of the resulting $W$-symmetry is beyond the scope of the
present work.
Likewise, the relevance of \wk to non-unitary $SU(2)$ coset models with
fractional level will not be considered here.
We only note that for fractional $k<-1$, there is no evidence that \wk
truncates to a finitely generated algebra and this might have implications to
non-unitary conformal field theories of fractional type (see for instance
\cite{26} and references therein).

(iv) \u{$Extension$ $of$ \wk $for$ $Rational$ $k$}: When $k$ is rational and
positive\footnote{For $k<0$ most of the parafermion operators have negative
spin.}, the chiral algebra of the theory is larger.
To see this, recall that the dimension of the parafermion  operator $\psi_{l}(
z)$ is $\Delta_{l}=l+l^{2}/k$.
Let us parametrize $k$ as $k=r/t$, where $r\in Z^{+}$, $t\in Z^{+}_{0}$ and
they are relatively prime.
It is obvious that the parafermion operators (and their adjoints) with $l=nr$,
$n\in Z$, have integer dimension $\Delta_{nr}=nr(1+nt)$.
Thus, we may consider a larger algebra, $\Psi^{r,t}_{\i}$,
which is an extension of \wk by the parafermion operators $\psi_{nr}$.
The complete description of a basis in this algebra is given by
$W^{n}_{s}(z)$ , where $n\in Z$ labels the $Z_{\i}$ charge sector (the
$Z_{\i}$ charge is $nr$) and $s=0,2,3,\cdots$. The dimension of $W^{n}_{s}(z)$
is $nr(1+nt)+s$.
$W^{0}_{s}(z)$ are the \wk generators with $W^{0}_{0}(z)=1$ and $W^{n}_{0}(z)
=\psi_{nr}(z)$.
The remaining generators are obtained from the parafermionic OPE and
the fusion rules implied by the $Z_{\i}$ symmetry are
$$[W^{n_{1}}_{s}]\otimes [W^{n_{2}}_{s'}]=\sum_{s''}[W^{n_{1}+n_{2}}_{s''}]
\;.\eqno(3.21)$$

When $r$ is even, we can define a local algebra even larger than $\Psi^{r,
t}_{\i}$, by including parafermions with half-integer spin as well.
For $r=2r'$, the parafermions $\psi_{nr'}$ have dimension $nr'(2+nt)/2$
and they form an algebra ${\tilde \Psi}^{2r',t}_{\i}$, where some of its
generators are fermionic.
Since we have assumed that $r$ and $t$ are relatively prime, $t$ is always odd
and $r'$ can be even or odd. If $r'$ is odd, then, $\psi_{nr'}$ with $n$ odd
will be fermionic. If $r'$ is even, then, the algebra
${\tilde \Psi}^{2r',t}_{\i}$ will contain
bosonic generators of integer spin only.
We also note that if $r$ is proportional to the square of an integer,
then $\Psi^{r,t}_{\i}$ can be extended further.
We will encounter this generalization later for $k=9/4$.

\bigskip
{\large\bf 4.} {\large\bf SL(2,R)/U(1) Characters}
\bigskip

In this section we are going to derive the characters for all unitary
\wk representations obtained from the \sl model.
Their derivation relies on the intimate relation between the $SL(2,R)$ current
algebra and the $N=2$ superconformal algebra with $c>3$, \cite{7}.
In section 2 we gave a brief description of this relation at the level of
the algebra.
As shown in \cite{7}, for any state in the base of an $SL(2,R)$
current algebra representation, there exists one highest weight (hw)
$N=2$ superconformal representation.
Let $\Phi_{m}^{j}(z)$ denote an operator\footnote{Only the chiral part is
important in this discussion.} of the $SL(2,R)$ WZW model with
$$J^{3}(z)\Phi_{m}^{j}(w)=m{\Phi_{m}^{j}(w)\over z-w}+{\cal O}(1)\,\,\,,\,\,\,
J^{\pm}(z)\Phi_{m}^{j}(w)=C_{\pm}(j,m){\Phi_{m\pm 1}^{j}(w)\over z-w}
+{\cal O}(1)\;,\eqno(4.1)$$
where $C_{\pm}^{2}=m(m\pm 1)-j(j-1)$.
Using eq. (2.10) we can factorize this operator as
$$\Phi_{m}^{j}=e^{m\sqrt{2\over k}\chi(z)}{\tilde \Phi}_{m}^{j}\;,\eqno(4.2)$$
where ${\tilde \Phi}_{m}^{j}$ is an operator in the \sl theory.
If we now define
$$Z_{m}^{j}=e^{-{2im\over k-2}\sqrt{(k-2)\over k}\varphi(z)}
{\tilde \Phi}_{m}^{j}(z)\;,\eqno(4.3)$$
then, $Z^{j}_{m}$ is an $N=2$ primary operator with dimension $\Delta$ and
$U(1)$ charge $Q$ given by
$$\Delta_{j,m} ={-j(j-1)+m^{2}\over (k-2)}\,\,\,,\,\,\,
Q_{j,m}=-{2m\over k-2}\,\,.\eqno(4.4)$$
\setcounter{footnote}{0}

The character of an $N=2$ hw representation $R$ is defined as
$\chi_{R}=Tr_{R}[q^{L_{0}}w^{J_{0}}]$. We are going to use the results of
\cite{27} where the characters of all unitary hw representations of the
$N=2$ superconformal algebra were derived from an analysis of the embedding
structure of the null Verma modules\footnote{The characters were also derived
independently in \cite{28} and for minimal representations in \cite{29}.}.

The unitarity restrictions on these representations (viewed either as \sl
or $N=2$ representations) are as follows: All the continuous series
representations are unitary. The trivial representation is also unitary.
{}From the $D^{\pm}_{n}$ representations only those satisfying $0\leq j\leq
k/2$
are unitary\footnote{An analysis of non-unitary \sl representations will be
presented elsewhere.}.
Thus, there are four different cases to consider corresponding to analogous
$SL(2,R)$ representations.

(i)  \u{$Trivial$ $Representation$}, $j=0,\, m=0$: This corresponds to the
vacuum representation of the $N=2$ algebra, $\Delta =0$, $Q=0$.
For $c>3$ there are two independent generating null vectors at relative charge
$\pm 1$ and level (mode number) $1/2$.
The character in this case is\footnote{It is enough for our purposes to
consider only the NS sector.}
$$\chi_{0}(q,w)=F_{NS}(q,w){1-q\over (1+q^{1/2}w)(1+q^{1/2}w^{-1})}\;,
\eqno(4.5)$$
where $F_{NS}$ is the unrestricted partition function
$$F_{NS}(q,w)=f(q)^{-2}\prod_{n=1}^{\i}(1+q^{n-{1\over 2}}w)
(1+q^{n-{1\over 2}}w^{-1})=f(q)^{-3}\sum_{n\in Z}q^{n^{2}/2}w^{n}
\eqno(4.6)$$
and
$$f(q)=\prod_{n=1}^{\i}(1-q^{n})\;.\eqno(4.7)$$

(ii) \u{$Positive$ $Discrete$ $Series$ ($D^{+}_{n}$)}:  For the positive
series we have
$j=n+\varepsilon$, where $0\leq \varepsilon <1$\footnote{$\varepsilon =0$
corresponds to $SO(2,1)$ representations, while $\varepsilon=0$ or
${1\over 2}$ to $SL(2,R)\sim SU(1,1)$ representations. A general $\varepsilon$
corresponds to the universal covering group.} and $m=j+r$, $r\in Z_{0}^{+}$.
These are lowest weight (lw) $SL(2,R)$ representations.
Here we have to distinguish two cases.

(iia) $j\not= k/2$: There is one generating null vector in the corresponding
$N=2$ representations at relative charge $-1$ and level $r+{1\over 2}$,
\cite{27}.
The character is
$$\chi^{+}_{j,m}(q,w)=q^{\Delta_{j,m}}w^{Q_{j,m}}{F_{NS}(q,w)\over
1+q^{r+{1\over 2}}w^{-1}}\;.\eqno(4.8)$$

(iib) $j=k/2$: In this case, apart from the charged null vector discussed
above,
there is also an independent one at relative charge zero and level 1.
Consequently, the character is
$$\chi^{+}_{{k\over 2},m}(q,w)=q^{\Delta_{k/2,m}}w^{Q_{k/2,m}}
{(1-q)F_{NS}(q,w)\over (
1+q^{r+{1\over 2}}w^{-1})(1+q^{r+{3\over 2}}w^{-1})}\;.\eqno(4.9)$$

(iii) \u{$Negative$ $Discrete$ $Series$ ($D^{-}_{n}$)}: For the negative series
we have $j=n-\varepsilon$, where $0\leq \varepsilon <1$ and  $m=-j-r$,
$r\in Z^{+}_{0}$.
These are hw $SL(2,R)$ representations and likewise we distinguish two cases.

(iiia) $j\not= k/2$: There is one generating null vector at level $r+{1\over
2}$ and relative charge 1. The character is
$$\chi^{-}_{j,m}(q,w)=q^{\Delta_{j,m}}w^{Q_{j,m}}{F_{NS}(q,w)\over
1+q^{r+{1\over 2}}w}\;.\eqno(4.10)$$

(iiib) $j=k/2$: In this case there is also an extra null vector at relative
charge zero and level 1. The character is
$$\chi^{-}_{{k\over 2},m}(q,w)=q^{\Delta_{k/2,m}}w^{Q_{k/2,m}}{(1-q)
F_{NS}(q,w)\over (1+q^{r+{1\over 2}}w)(1+q^{r+{3\over 2}}w)}\;.\eqno(4.11)$$

(iv) \u{$Continuous$ $Representations$}: These correspond to the principal and
complementary series of $SL(2,R)$ representations. In this case $j(j-1)
<\varepsilon(\varepsilon-1)$ and the representations are neither hw nor lw.
In the principal series, $j={1\over 2}+i\rho$, $\rho\in R$, $m=m_{0}+n$,
$n\in Z$, $|m_{0}|\leq 1/2$.
In the supplementary series $j$ is real with $0<|j-{1\over
2}|<\varepsilon+{1\over 2}$, $m=m_{0}+n$, $n\in Z$, $|m_{0}\pm{1\over
2}|<|j-{1\over 2}|$.
These give rise to $N=2$ representations with irreducible Verma modules.
Consequently, the character is
$$\chi^{c}_{j,m}(q,w)=q^{\Delta_{j,m}}w^{Q_{j,m}}F_{NS}(q,w)\;.\eqno(4.12)$$

To go from $N=2$ characters to \sl characters is rather straightforward.
At the representation level, one has to decompose the $N=2$ representations
into
$U(1)$ representations and keep only the hw state of each $U(1)$
representation. This is done by expanding the $N=2$ character in a power series
in $w$. The contribution of any given $U(1)$ hw representation of charge $Q$
appears as a factor multiplying $w^{Q}$. To factor out the $U(1)$
representation, but its hw state, we must divide by the $U(1)$ multiplicity
function $f(q)$.
One also needs to correct the dimension of a given $U(1)$ hw state
of charge $Q$ by subtracting a factor ${k-2\over 2k}Q^{2}$.
This way we can obtain the characters of the parafermionic \sl model,
$\psi_{j,m}(q)$, which define the non-compact analogue of the string
functions $\cc^{j}_{m}(q)$ via
$$\psi_{j,m}(q)=f(q)\,\,\cc^{j}_{m}(q)\;,\eqno(4.13)$$
where $f(q)$ was defined by eq. (4.7).
Another point to be stressed is the following. In the $N=2$ case, starting
from a hw state with a certain $U(1)$ charge and acting with the
supercharges, one moves to states that correspond to $SL(2,R)$ states with
different $J^{3}_{0}$ eigenvalue compared to the initial one.
Thus, for fixed $j$, starting with a certain $m$ and decomposing the $N=2$
character, we can obtain the characters of the whole coset module generated
by the representation labeled by $j$.

We summarize the results for the non-compact string functions.\footnote{The
characters in cases (iia) and (iii) have been calculated previously in
\cite{30}.}

(i) \u{$Trivial$ $Representation$}:
$$\cc^{j=0}_{m}(q)={q^{|m|+{m^{2}\over k}}\over
f(q)^{3}}\left[1+\sum_{n=1}^{\i}
(-1)^{n}q^{{n^{2}+(2|m|+1)n-2|m|\over 2}}(1+q^{|m|})\right]\,\,,\,m\in Z\;.
\eqno(4.14)$$

(iia) \u{$D^{\pm}_{n}$}, $j\not= k/2$:
$$\cc^{j}_{m}(D;q)={q^{-{j(j-1)\over k-2}+{(j+m)^{2}\over k}}\over f(q)
^{3}}\sum_{n=0}^{\i}(-1)^{n}q^{n(n+2m+1)\over 2}\;,\,\,\,m\in Z\;.\eqno(4.15)$$

(iib) \u{$D^{\pm}_{n}$}, $j= k/2$:
$$\cc^{k/2}_{m}(D;q)=\cc^{j=0}_{m}(q)\;.\eqno(4.16)$$

(iii) \u{$Continuous$ $Representations$}:
$$\cc^{j}_{m}(c;q)={z^{-{j(j-1)\over k-2}+{m^{2}\over k}}\over f(q)^{3}}\,\,\,,
\,\,\,m=m_{0}+n\,\,\,,\,\,\,n\in Z\;.\eqno(4.17)$$
We should remark that the identity (4.16) also holds in the compact $SU(2)$
case.

Using the above we can write the following decomposition formulae for the $N=2$
characters, which are generalizations of the compact case, \cite{31}.
For the $D^{\pm}$ representations we obtain
$$\chi^{j}_{m=\pm(j+r)}(q,w)=\sum_{n\in Z}q^{{k-2\over 2k}
\left(n-r-{2(j+r)\over k-2}\right)^{2}}w^{n-r-{2(j+r)\over k-2}}\,\,
\cc^{j}_{n}(D;q)\;\,\,,
r\in Z^{\pm}_{0}\eqno(4.18)$$
respectively.
For $j=0$ we obtain the result for the trivial representation,
while for the continuous series,
$$\chi^{j}_{m}(q,w)=\sum_{n\in Z}q^{{k-2\over 2k}\left(n-{mk\over k-2}
\right)^{2}}w^{n-{2m\over k-2}}\,\,\cc^{j}_{m-n}(c;q)\;.\eqno(4.19)$$
The relations above differ from the compact case ($N=2$ minimal models)
, where the string functions are periodic
with period $2k$, due to the $Z_{k}$ symmetry . This turns the infinite sum
into a $mod$ $2k$ sum and
produces the $SU(2)$ $\vartheta$-functions, \cite{31,32},
$$\chi_{j,m}^{\rm compact}=\sum_{m'=-k+1}^{k}\,c_{m'}^{j}(q)
\vartheta_{m'(k+2)-mk,k(k+2)}\left({\tau\over 2},{z\over k+2}\right)\;,
\eqno(4.20)$$
where
$$\vartheta_{m,k}(\tau,z)=\sum_{n\in Z+m/2k}q^{kn^{2}}w^{kn}\,\,\,,\,
q=e^{2\pi i\tau}\,\,,\,\,w=e^{2\pi i z}\eqno(4.21)$$
is the standard $\vartheta$-function.

We can now compute the characters of all $SL(2,R)_{k}$
representations which give rise to unitary \sl representations.
We summarize the formulae for the affine characters, defined as
$Tr[q^{L_{0}}w^{J_{0}^{3}}]$.

(i) \u{$Trivial$ $Representation$}:
$$\chi_{0}^{SL_{2}}(q,w)={1\over \pi(q,w)}\;,\eqno(4.22)$$
where
$$\pi(q,w)=\prod_{n=1}^{\i}(1-q^{n})(1-q^{n}w)(1-q^{n}w^{-1})\;.\eqno(4.23)$$

(ii) \u{$D^{+}$ $Representations$}, $j\not= k/2$:
$$\chi_{j}^{SL_{2}^{+}}(q,w)={q^{-{j(j-1)\over k-2}}w^{j}\over (1-w)\pi(q,w)
}\;.\eqno(4.24)$$
For the $D^{+}$ representation with $j=k/2$, the Verma module is reducible
because there is a null vector at level one.
This is the only hw representation embedded in it, since it transforms as a
$j'={k\over 2}-1$ representation.
Consequently, the character is
$$\chi_{k/2}^{SL_{2}^{+}}(q,w)={q^{-{j(j-1)\over k-2}}w^{j}(1-qw^{-1})\over (
1-w)\pi(q,w)}\;.\eqno(4.25)$$

(iii) \u{$D^{-}$ $Representations$}:
$$\chi_{j}^{SL_{2}^{-}}(q,w)={q^{-{j(j-1)\over k-2}}w^{-j}\over (1-w^{-1})
\pi(q,w)}\;,\eqno(4.26)$$
$$\chi_{k/2}^{SL_{2}^{-}}(q,w)={q^{-{j(j-1)\over k-2}}w^{-j}(1-qw)\over
(1-w^{-1})\pi(q,w)}\;.\eqno(4.27)$$
For the continuous series the characters converge nowhere in the complex
$w$-plane.

The $SL(2,R)_{k}$ characters above are related to the non-compact string
functions via a generalization of the Ka\v c-Peterson  formula, \cite{33},
$$\chi^{SL_{2}^{\pm}}_{j}(q,w)=\sum_{m\in Z}q^{-{(j+m)^{2}\over k}}w^{\pm
j+m}\,{\hat c}^{j}_{m}(D;q)\;.\eqno(4.28)$$
\setcounter{footnote}{0}

The string functions ${\hat c}^{j}_{m}$ are the characters of \wk
representations with $Z_{\i}$ charge $m$\footnote{Sometimes string functions
are referred to as parafermionic characters.
This is not correct.
A parafermionic representation whose module is generated by the action of the
parafermionic currents decomposes into a finite sum of $W$-representations in
the $SU(2)$ case and into an infinite sum of \wk representations in the
$SL(2,R)$ case.
Parafermionic representations are in one to one correspondence with
$SL(2,R)$ representations.}.
In particular, the string functions associated with the trivial ($j=0$)
representation are the \wk characters of the parafermions $\psi_{m}(z)$, with
$m>0$ and their adjoints, with $m<0$.
${\hat c}^{0}_{0}$ is the character of the vacuum module of the \wk algebra.
It is obvious that the state multiplicities in all \wk characters are
independent of $k$.
Consequently, the counting of independent states, null states and generators is
the same for all $k>2$ and in particular, it coincides with that of $k=\i$.
The number and structure of null states of \wk representations can be obtained
from the explicit realization of \w in \cite{5}.
{}From eq. (4.28), we can obtain for $j=0$ the non-trivial identity
$$\sum_{m\in Z}{\hat c}^{0}_{m}(q)=f(q)^{-3}+{\cal O}(1/k)\;,\eqno(4.29)$$
which describes the decomposition of the vacuum representation of the $U(1)
\times U(1)$ current algebra into \w representations, \cite{34}.

\bigskip
{\large\bf 5.} {\large\bf Applications to 2-d Quantum Gravity}
\bigskip

The presence of the \wk symmetry in the \sl coset model suggests the existence
of an infinite number of conservation laws for all $k\geq 2$ and in particular
for $k=9/4$, which corresponds to Witten's black hole solution.
Let us introduce the following charges
$$Q_{s}=\oint dz\,W_{s}(z)\; ,\;\;s\geq 2\,\,,\eqno(5.1)$$
which by definition are identified with the $-s+1$ Fourier mode of the
$W_{s}(z)$ fields, i.e.,
$$Q_{s}=W_{s}^{(-s+1)}\,\,\,.\eqno(5.2)$$
Clearly, in order to insure that all $Q_{s}$ are in involution,
$$[Q_{s},Q_{s'}]=0\;,\;\;\forall\,s,s'\geq 2\;,\eqno(5.3)$$
the $W$-generators of the algebra must satisfy the condition
$$\oint\,dz\,\oint\,dw\,W_{s}(z)W_{s'}(w)=0\;.\eqno(5.4)$$

At this point one might think that the non-linear terms of the \wk algebra
violate the condition (5.4).
This condition is certainly valid in the limit $k\rightarrow\i$,
after introducing the standard quasiprimary field basis of \w, \cite{4,5}.
Eq. (5.4) will be true in general, iff the coefficient of $(z-w)^{-1}$ in the
OPE $W_{s}(z)W_{s'}(w)$ is a total derivative.
It follows from the results we have already presented that the OPE for $s=3$,
$s'=4$ does not enjoy the desired property.
Hence, the question arises whether there are appropriate redefinitions of the
$W$-generators which can give rise to infinitely many charges in involution.
Our attitude is that this is indeed the case, although we cannot provide a
complete proof because the commutation relations of \wk are not yet known in
closed form.
Later we will reformulate this conjecture in a language which is analogous to
the conservation laws of the KP hierarchy and suggest an alternative
way for interpreting the charges $Q_{s}$ in involution.
We also note that the field redefinitions we are searching for $s\geq 4$ should
yield the quasiprimary field basis of \w in the limit $k\rightarrow \i$, for
which eqs. (5.3) are automatic.

Let us now sketch the basic steps for constructing a new spin-4 generator which
is compatible with eq. (5.4).
Apart from the OPE (3.9) we also have
$$W_{3}(z+\e):W_{2}^{2}:(z)={48\over 5}\left({1\over \e^{4}}+{1\over
3}{\pa\over \e^{3}}+{1\over 14}{\pa^{2}\over \e^{2}}+{1\over
84}{\pa^{3}\over \e}\right)W_{3}(z)+$$
$$+6\left({1\over \e^{2}}+{2\over 5}{\pa\over \e}\right):W_{2}W_{3}:(z)+4{:(\pa
W_{2})W_{3}:(z)\over \e}+{\cal O}(1)\,\,,\,\,\eqno(5.5)$$
which can be computed directly, using the free field realization of the
$W$-generators, or more easily using the general formulae
$${\langle :W_{2}^{2}:(z_{1})W_{s}(z_{2})W_{s}(z_{3})\rangle\over
\langle W_{s}(z_{2})W_{s}(z_{3})\rangle}=s(s+{1\over 5})
{z_{23}^{4}\over z_{12}^{4}z_{13}^{4}}\;,\eqno(5.6a)$$
$${\langle :W_{2}W_{s}:(z_{1})W_{s}(z_{2}):W_{2}^{2}:(z_{3})\rangle\over
\langle W_{s}(z_{1})W_{s}(z_{2})\rangle}=s\left[c+{2s(8s-5)\over 2s+1}\right]
{z_{12}^{2}\over z_{13}^{6}z_{23}^{2}}\;,\eqno(5.6b)$$
$${\langle :(\pa W_{2})W_{s}:(z_{1})W_{s}(z_{2}):W_{2}^{2}:(z_{3})\rangle\over
\langle W_{s}(z_{1})W_{s}(z_{2})\rangle}
={4s(s-1)\over s+2}\left[c+{(s-2)(3s-1)\over s+1}\right]
 {z_{12}\over z_{13}^{7}z_{23}}\;.\eqno(5.6c)$$
Here $c$ is the central charge of the Virasoro algebra and $W_{s}$ are Virasoro
primaries.
Then, the combination
$$W^{\rm new}_{4}(z)=W_{4}(z)+48{(k+3)(2k-3)(k-2)\over k^{2}(16k-17)}:
W_{2}^{2}:(z)\eqno(5.7)$$
has an OPE with $W_{2},W_{3}$ with single pole that contains only total
derivatives, as required.

We point out that the spin 4 field (5.7) is different from ${\tilde W}_{4}(z)$
introduced earlier (cf. eq. (3.11)).
They both coincide, however, in the the limit $k\rightarrow \i$, according to
the general expectation we mentioned.
Carrying out this procedure for other higher spin fields is straightforward,
but cumbersome and it will not be dealt with here.
It would be interesting to describe the complete structure of \wk,
not in terms of primary field generators, but in a basis which makes the
conditions (5.4) manifest.
Whether this is possible in practice or a formidable calculational task
remains to be seen.

Before we proceed further, two remarks are in order, with emphasis on Witten's
black hole solution.
First, for $k=9/4$, the parafermion fields $\psi_{l}(z)$ with $l=3n$, $n\in Z$,
are local operators with integer dimension
$$\Delta_{l}=n(4n+3)\,\,\,.\eqno(5.8)$$
Therefore, as we explained earlier, one might consider enlarging \wk by
including $\{ \psi_{3n}(z)\}$ in the spectrum.
This extended $W$-algebra governs the physics of the 2-d
black hole solution and we expect it to be the maximal symmetry of the model.
It would be very interesting to analyze its structure in detail and study the
algebra of charges associated with the $\psi_{3n}$ generators.
Of course, similar questions can be raised for arbitrary (but rational) values
of $k$.
As for the 2-d black hole, it would be interesting to know whether there is any
relation between the parafermionic extension we are
considering here and the algebra of volume preserving diffeomorphisms of the
3-d cone introduced by Witten in his recent study of 2-d string theory,
\cite{35}.

Second, the existence and explicit construction of an infinite family of
independent charges in involution is very important for understanding the
physics of \sl coset models, in general.
For the black hole solution, in particular, the set $\{Q_{s}\}$ could be
responsible for the maintenance of quantum coherence in the Hawking evaporation
process (see for instance \cite{36} for some preliminary results in this
direction).
It is an interesting problem to find the target space interpretation of the
\wk symmetry (and its parafermionic extensions for fractional
$k$) and examine whether the quantum mechanical states of the theory can be
completely classified by the corresponding conserved charges.
We hope to address these issues elsewhere.

Next, we present some thoughts
aiming at a deeper connection between the conservation laws of the \sl
coset and those of the KP hierarchy.
The basic idea here originates in the recent works of Yu and Wu, \cite{17,19},
who studied the bi-Hamiltonian structure of the KP hierarchy (see also
\cite{18,20}).
We review their main results for the completeness of our presentation.
The interpretation we will put forward for the charges $Q_{s}$ is essentially
a quantum mechanical version of the classical KP integrals.

Recall that the KP hierarchy can be formulated as a Lax pair, using
the pseudo-differential operator
$$L=\pa_{z}+\sum_{r=0}^{\i}u_{r}(z)\pa^{-r-1}_{z}\eqno(5.9)$$
(for details see \cite{37} and references therein).
For the Hamiltonian description of the KP hierarchy we introduce the following
quantities
$$H_{s}\equiv {1\over s}{\rm res}\,L^{s}\; ,\;\;s\,\in\,Z^{+}\eqno(5.10)$$
and consider the corresponding charges,
$${\cal H}_{s}=\oint\,dz\,H_{s}(z)\;,\eqno(5.11)$$
as Hamiltonian functionals for the various flows.

The first Hamiltonian structure of the KP hierarchy is given by the commutation
relations of \ww with zero central charge, which in terms of the $\{u_{r}(z)\}$
variables assume the Watanabe form, \cite{38},
$$\{u_{r}(z),u_{s}(w)\}_{(1)}=K^{(1)}_{rs}(z)\delta
(z-w)\,\,\,\,,\eqno(5.12a)$$
$$K^{(1)}_{rs}(z)=\sum_{l=0}^{r}(-1)^{l}{r \choose l}u_{r+s-l}(z)\pa_{z}^{l}-
\sum_{l=0}^{s}{s \choose l}\pa_{z}^{l}u_{r+s-l}(z)\,\,.\eqno(5.12b)$$
The appropriate change of variables that establishes the isomorphism with
\ww is given by
$$V_{s}(z)={(s-1)!\over 2^{s-1}(2s-3)!!}\sum_{l=0}^{s-1}{s-1\choose
l}{2s-l-2\choose s-1}u^{(l)}_{s-l-1}(z)\,\,\,.\eqno(5.13)$$
It also follows from the integrability properties of the KP hierarchy that the
charges (5.11) are in involution, i.e.,
$$\{{\cal H}_{s},{\cal H}_{s'}\}_{(1)}=0\;\;.\eqno(5.14)$$
If we change basis from (5.13) to (5.10), the commutation relations of \ww
will assume a non-linear form, which Yu and Wu related to the second
Hamiltonian
structure of the KP hierarchy as follows,
$$\oint \,dw\,\{H_{s}(z),H_{s'}(w)\}_{(2)}=\oint \,dw\,\{H_{s}(z),
H_{s'+1}(w)\}_{(1)}\;.\eqno(5.15)$$

The second structure of the KP hierarchy is non-linear, given by the
commutation relations
$$\{u_{r}(z),u_{s}(w)\}_{(2)}=K^{(2)}_{rs}(z)\delta (z-w)\,\,\,,\eqno(5.16a)$$
where
$$\eqalign{\phantom{K^{(2)}_{rs}(z)}K^{(2)}_{rs}(z)&=\sum_{l=0}^{r+1}(-1)^{l}
{r+1\choose l}u_{r+s+1-l}
(z)\pa_{z}^{l}-\sum_{l=0}^{s+1}{s+1\choose l}\pa_{z}^{l}u_{s+r+1-l}(z)+\cr
&+\sum_{l=0}^{r-1}\sum_{k=0}^{s-1}(-1)^{r+l}{r\choose l}{s\choose k}u_{l}(z)
\pa_{z}^{r+s-l-k-1}u_{k}(z)+\cr
&+\sum_{l=0}^{\i}\Biggl[\sum_{m=r+1}^{r+s}\sum_{k=0}^{m-r-1}(-1)^{l+k}
{m-k-1\choose l-k}{s\choose k}-\cr
&-\sum_{m=l+1}^{l+s}(-1)^{l}{m-s-1\choose l}\Biggr]
u_{m-l-1}(z)\pa_{z}^{l}u_{r+s-m}(z)\;\;.\cr}\eqno(5.16b)$$
Change of basis to the variables (5.10) maintains the non-linear features of
(5.16) and gives rise to a non-linear deformation of the centerless \w,
denoted by ${\hat W}_{\i}$ in \cite{19}.

The ${\hat W}_{\i}$ algebra of Yu and Wu is classical with zero central
charge, but it resembles the general structure of \wk, provided that
$H_{s}(z)$ is identified with $W_{s+1}(z)$ ($s\geq 1$).
For comparison we also have to introduce appropriate redefinitions which
transform \wk to a non-primary field basis.
The recursion relations (5.15) are sufficient to insure that the charges
$Q_{s+1}\equiv  {\cal H}_{s}$ of ${\hat W}_{\i}$ are all in involution.
This follows by integrating eq. (5.15) with respect to $z$ and noting that its
right hand side subsequently vanishes, thanks to eq. (5.14).
Therefore, although the relation between the classical algebras \ww and
${\hat W}_{\i}$ is indirect and valid only in the integrated form of the
commutation relations (5.15), the implications for ${\hat W}_{\i}$ are
substantial and not at all obvious without the KP point of view.
The most important point for us is the identification of the ${\hat W}_{\i}$
charges with the conserved quantities of the KP hierarchy and their explicit
construction in terms of \ww generators, as dictated by eq. (5.13).

We are now in the position to state our proposal for the conservation laws
of the 2-d black hole solution, after choosing a suitable non-primary
basis for the generators of \wk.
{\em The charges $Q_{s}$ are the conserved quantities of a quantum deformation
of the KP hierarchy.}
In the theory of quantum integrable systems, usually only one of the two
Hamiltonian structures survives, \cite{39} and therefore recursive relations
of the form (5.15) cannot be used to prove complete integrability.
The latter is established by explicit computation case by case,
which is a very cumbersome procedure, as the one we encounter for the black
hole.
Quantum equations of KdV type are naturally defined by means of their quantum
second Hamiltonian structure.
The obvious candidate for our problem is \wk with $c_{k}\not=0$ and this is
the essence of the proposal.

The quantum deformation of the KP hierarchy we advocate for interpreting
the black hole charges is rather speculative at this point and further work
is required to illuminate it.
If it exists, however, it would have important consequences on the relation
between \sl coset models and the ordinary KP approach to non-perturbative
string theory via matrix models.
In this framework, it will be also interesting to examine whether recursive
relations between \wk and $W_{1+\i}$ with non-zero central charge still exist
for some miraculous reason, for certain values of $k$.
We also hope to address these issues elsewhere.

\bigskip
{\large\bf 6.} {\large\bf Conclusions and Further Remarks}
\bigskip

In this paper we studied the generic chiral symmetry of the \sl
coset model.
It is described by the non-linear algebra \wk with an infinite number of
generators of spin $s=2,3,\cdots$.
This is a universal  $W$-algebra (for the $A$-series of extended conformal
symmetries), since at $k=-N$ with $N=1,2,\cdots$, it
truncates to the usual Zamolodchikov $W_{N}$ algebras.
As $k\rightarrow\pm \i$, one recovers the usual \w algebra, which is linear
in a quasiprimary field basis.
We have investigated the general structure of \wk, using its relation to
compact and non-compact parafermions.
We have also constructed a free field realization of the algebra in terms of
two bosons with background charge.
For $k=2$, the central charge of the Virasoro subalgebra is infinite,
but as we argued, the generators with spin $s\geq 3$ form a well-defined closed
linear subalgebra.
Several other truncations seem to be possible for certain special values of
$k<2$.
For rational values of $k=r/t$, the chiral algebra \wk can be enlarged further
by including all parafermionic operators with integer (or even half-integer)
spin.
This yields a double graded algebra which definitely deserves more attention,
in particular for $k=9/4$.
The complete structure of \wk in closed form remains an open problem.

The spectrum of the \sl coset model is built out of hw representations
of \wk.
We computed the characters (generalized string functions) of all unitary
representations of \wk, using the relationship between \sl and the $N=2$
superconformal representation theory with $c>3$.
The characters for the vacuum module provided crucial information on the
spectrum of \wk, which is the same and infinitely generated for all $k>2$.
As a byproduct, we also obtained the $SL(2,R)_{k}$ characters for
representations in the discrete series.

\wk is related to various approaches to 2-d gravity in two possible
ways. First, it is a symmetry of the black hole solution (for $k=9/4$).
The existence of an infinite number of integrals in involution will
characterize
the structure of string theory in the black hole background.
This could have important implications on questions of loss of coherence
during black hole evaporation.
We conjectured the existence of such integrals in involution,
based on explicit calculations up to $s=4$ and their existence for all $s$
in the large $k$ limit.
This algebra also seems to be a quantum version of the second Hamiltonian
structure of the KP hierarchy.
Due to the importance of the this hierarchy to matrix model approaches
to 2-d quantum gravity, it is reasonable to expect that \wk could help
illuminating the KP-like integrable structure of 2-d string theory.

Finally, we conclude by discussing some further generalizations of the
$W$-symmetry of \sl to higher rank coset models.
In \cite{5} we constructed a ``colored" generalization of the usual \w
algebra, denoted by $W^{p}_{\i}$, which is linear and contains \w as a
subalgebra.
$W^{p}_{\i}$ arises as the $N\rightarrow \i$ limit of the chiral algebra
of the Grassmannian coset models $SU(p+1)_{N}/(SU(p)_{N}\times U(1))$.
One may consider the non-compact version of these models,
$SU(p,1)_{k}/(SU(p)_{k}\times U(1))$ and examine whether the corresponding
non-linear chiral algebra has the same spectrum as $W^{p}_{\i}$.
Again, when $k$ becomes a negative integer, the algebra should truncate to the
finitely generated chiral algebra of the compact coset.
It would be interesting to investigate this generalization of \wk, since it
also qualifies as a universal $W$-algebra for the $A$-series of extended
conformal symmetries, but for a wider range of values for the central charge.
It also remains a challenging question to find a framework for constructing
the most universal $W$-symmetry that contains all possible chiral algebras
of 2-d CFT.
Such master symmetry could play a prominent role in the non-perturbative study
of string theory.

\bigskip
\centerline{\bf Acknowledgements}

This work was initiated while both of us were attending the
``{\em Infinite Analysis}" workshop at the Research Institute for
Mathematical Sciences in Kyoto.
We are grateful to Prof. T. Miwa and all other members of RIMS for their kind
invitation, generous financial support and Japanese-style hospitality, which
provided us with a very enjoyable and stimulating environment.
Special thanks are due to Prof. T. Inami for the hospitality extended to us at
the Yukawa Institute for Fundamental Physics.
We are also grateful to Prof. L. Alvarez-Gaum\'e and all other members of the
Theory Division at CERN for their hospitality and financial support during the
last stages of this work. E. K. would also like to thank the organizers of the
Aspen Summer Institute and the Ecole Normale Summer Institute
for hospitality and financial support, during the course of this work.

\bigskip
\centerline{\bf Note Added}

After the completion of this work we received ref. \cite{40}, where
the parafermionic characters discussed in section 4 were independently
obtained.
\newpage
\centerline{\bf Appendix A}

In this appendix we give the free field realization of the primary
$W_{4}(z)$ and $W_{5}(z)$ generators.
Normal ordering is implicitly assumed.

$$\eqalign{\phantom{4}W_{4}(z)&=-{4\over k^{2}(16k-17)}\Bigl[ \a_{1}
(\pa\pt)^{4}+\a_{2}
(\pa\pp)^{4}+\a_{3}(\pa\pt)^{2}(\pa\pp)^{2}+\a_{4}(\pa^{2}\pp)^{2}+\cr
&+\a_{5}(\pa^{2}\pt)^{2}+\a_{6}\,\pa\pp\pa^{3}\pp+\a_{7}\,\pa\pt\pa^{3}\pt
+\a_{8}\,\pa^{4}\pt+\a_{9}\,\pa^{2}\pt(\pa\pp)^{2}+\cr
&+\a_{10}\,\pa^{2}\pt(\pa\pt)^{2}+\a_{11}\,\pa\pt\pa\pp\pa^{2}\pp\Bigr]\;,
\cr}\eqno(A.1)$$
where
$$\matrix{\a_{1}=(k-2)^{2}(6k+5)\hfill&,&\a_{2}=(k-12)(2k-1)(3k-4),\hfill\cr
\a_{3}=6(k-2)(2k^{2}-13k+8)\hfill&,&\a_{4}=-2(3k-4)(2k^{2}+2k+3),\hfill\cr
\a_{5}=-2(k-2)(6k^{2}-12k+1)\hfill&,&\a_{6}=\a_{7}=8(k-2)(k^{2}-k+1),\hfill\cr
\a_{8}=-{2\over 3}(k^{2}-k+1)\kl\hfill&,&\a_{9}=-2(2k-3)(19k-8)\kl,\hfill\cr
\a_{10}=-2(k-2)(6k+5)\kl\hfill&,&\a_{11}=4k(16k-17)\kl.\hfill\cr}\eqno(A.2)$$

$$\eqalign{\phantom{W_{5}(}W_{5}(z)&={16i\over 5k^{2}(64k-107)}\k
\Biggl[\b_{1}(\pa\pp)^{5}
+\b_{2}(\pa\pt)^{4}\pa\pp+\b_{3}(\pa\pt)^{2}(\pa\pp)^{3}+\cr
&+\b_{4}(\pa\pt)^{2}\pa^{3}\pp+\b_{5}(\pa\pp)^{2}\pa^{3}\pp+\b_{6}\,
\pa\pp(\pa^{2}\pp)^{2}+\b_{7}\,\pa^{5}\pp+\cr
&+\b_{8}\,\pa\pt\pa^{2}\pt\pa^{2}
\pp+\b_{9}(\pa^{2}\pt)^{2}\pa\pp+\b_{10}\,\pa\pt\pa^{3}\pt\pa\pp+\cr
&+\kl\Bigl(\gg_{1}(\pa\pt)^{3}\pa^{2}\pp+\gg_{2}\,\pa^{2}\pt(\pa\pp)^{3}+
\gg_{3}(\pa\pt)^{2}\pa^{2}\pt\pa\pp+\cr
&+\gg_{4}\,\pa\pt(\pa\pp)^{2}\pa^{2}\pp+\gg_{5}\,\pa^{3}\pt\pa^{2}\pp+
\gg_{6}\,\pa^{2}\pt\pa^{3}\pp+\cr
&+\gg_{7}\,\pa^{4}\pt\pa\pp+\gg_{8}\,\pa\pt\pa^{4}\pp\Bigr)\Biggr]\;,\cr}
\eqno(A.3)$$
where
$$\matrix{\b_{1}=6(2k-1)(5k-24)(5k-8)\hfill&,&\b_{2}=30(k-2)^{2}(10k+7),
\hfill\cr
\b_{3}=60(k-2)(2k-3)(5k-8)\hfill&,&\b_{4}=-15(k-2)(2k^{2}-7k-16),\hfill\cr
\b_{5}=15(5k-8)(6k^{2}-7k+8)\hfill&,&\b_{6}=-30(5k-8)(4k^{2}+k+6),\hfill\cr
\b_{7}=k(k^{2}-3k+5)\hfill&,&\b_{8}=180(k-2)(3k^{2}-4),\hfill\cr
\b_{9}=-30(k-2)(38k^{2}-63k-2)\hfill&,&\b_{10}=30(k-2)(16k^{2}-31k+16),
\hfill\cr
\gg_{1}=-15k(k-2)(10k+7)\hfill&,&\gg_{2}=15(5k-8)(2k^{2}-33k+12),\hfill\cr
\gg_{3}=15(k-2)(k-4)(10k+7)\hfill&,&\gg_{4}=-15k(5k-8)(2k-25),\hfill\cr
\gg_{5}=-15(2k-3)(2k^{2}+k+8)\hfill&,&\gg_{6}=15(4k^{3}-10k^{2}
+13k-16),\hfill\cr
\gg_{7}=10(k-2)(k^{2}-3k+5)\hfill&,&\gg_{8}=-10k(k^{2}-3k+5).\hfill\cr}
\eqno(A.4)$$
\newpage
\vskip .6in
\centerline{\bf Appendix B}

In this appendix we give the free field realization of the composite
operators appearing in the OPEs (3.7) and (3.9).
\bigskip
$$\eqalign{:W^{2}_{2}:(z)&={1\over 4}(\pa\pt)^{4}+{1\over 4}(\pa\pp)
^{4}-{1\over 5}\,\pa\pp\pa^{3}\pp-{1\over 5}\,\pa\pt\pa^{3}\pt+\cr
&+{3\over 10}(\pa^{2}\pp)^{2}+{3k-1\over 10(k-2)}(\pa^{2}\pt)^{2}
+{1\over 2}(\pa\pt)^{2}(\pa\pp)^{2}+\cr
&+{1\over \kl}\Bigl({1\over 30}\,\pa^{4}\pt
-\pa^{2}\pt(\pa\pp)^{2}-(\pa\pt)^{2}\pa^{2}\pt\Bigr)\;,\cr}\eqno(B.1)$$
\bigskip
$$\eqalign{\phantom{W}:W_{2}W_{3}:(z)&={i\over 84k}\k\Biggl[-28(3k-4)
(\pa\pp)^{5}-84(k-2)(\pa\pt)^{4}\pa\pp-\cr
&-56(3k-5)(\pa\pt)^{2}(\pa\pp)^{3}+2(17k-18)(\pa\pt)^{2}\pa^{3}\pp-\cr
&-12(19k-24)\,\pa\pt\pa^{2}\pt\pa^{2}\pp+2(65k-96)(\pa\pp)^{2}\pa^{3}\pp-\cr
&-72(3k-4)\pa\pp(\pa^{2}\pp)^{2}+12(k-2)(\pa^{2}\pt)^{2}\pa\pp+\cr
&+k\,\pa^{5}\pp+96(k-2)\,\pa\pt\pa^{3}\pt\pa\pp\Biggr]+\cr
&+{i\over 21k\sqrt{k(k-2)}}\Biggl[-7(3k^{2}-18k+20)\,\pa^{2}\pt(\pa\pp)^{3}+\cr
&+21k(k-2)(\pa\pt)^{3}\pa^{2}\pp+21k(k-2)\pa\pt(\pa\pp)^{2}\pa^{2}\pp-\cr
&-21(k-2)(k-4)(\pa\pt)^{2}\pa^{2}\pt\pa\pp+5(k-2)^{2}\,\pa^{4}\pt\pa\pp-\cr
&-5k(k-2)\pa\pt\pa^{4}\pp-6(k-2)(5k-6)\,\pa^{3}\pt\pa^{2}\pp+\cr
&+(30k^{2}-77k+48)\,\pa^{2}\pt\pa^{3}\pp\Biggr]\;,\cr}\eqno(B.2)$$

$$\eqalign{:(\pa W_{2})W_{3}:(z)&={i\over
105k}\k\Biggl[7(k-2)\Bigl(\pa\pt\pa^{4}
\pt\pa\pp-6(\pa\pp)^{2}\pa^{4}\pp-\cr
&-6(\pa\pt)^{3}\pa^{2}\pt\pa\pp+6(\pa\pt)^{4}\pa^{2}\pp\Bigr)+
(3k-4)\Bigl(-15(\pa^{2}\pp)^{3}-\cr
&-42\,\pa\pt\pa^{2}\pt(\pa\pp)^{3}+42(\pa\pt)^{2}(\pa\pp)^{2}\pa^{2}\pp
-93\,\pa\pp\pa^{2}\pp
\pa^{3}\pp\Bigr)-\cr
&-(13k-66)\,\pa\pt\pa^{2}\pt\pa^{3}\pp+{7\over 2}(3k-2)(\pa\pt)
^{2}\pa^{4}\pp-\cr
&-114(k-1)\,\pa\pt\pa^{3}\pt\pa^{2}\pp+3(11k+10)(\pa^{2}\pt)^{2}\pa^{2}
\pp\Biggr]-\cr
&-{i\over 15k\sqrt{k(k-2)}}\Biggl[(k-2)\Bigl(\pa^{4}\pt\pa^{2}\pp-18\pa^{3}\pt(
\pa\pt)^{2}\pa\pp+\cr
&+12\pa^{2}\pt(\pa\pt)^{2}\pa^{2}\pp+24\pa\pt(\pa^{2}\pt)^{2}\pa\pp\Bigr)+\cr
&+6(3k-4)\Bigl(2\pa^{2}\pt(\pa\pp)^{2}\pa^{2}\pp-\pa^{3}\pt(\pa\pp)^{3}\Bigr)
+\cr
&+6(k^{2}-5k+5)\pa^{3}\pt\pa^{3}\pp+
(6k^{2}-21k+22)\pa^{2}\pt\pa^{4}\pp\Biggr]
\;.\cr}\eqno(B.3)$$
\vskip .2in
\centerline{\bf Appendix C}

In this appendix we give the two-point functions for the first few $W$-fields,
normalized as in eq. (3.4).

$$\langle W_{2}(z)W_{2}(w)\rangle={k+1\over k-2}{1\over (z-w)^{4}}\;,
\eqno(C.1)$$
$$\langle W_{3}(z)W_{3}(w)\rangle={16\over 3}{(k+1)(k+2)(3k-4)\over
k^{3}}{1\over (z-w)^{6}}\;,\eqno(C.2)$$
$$\langle W_{4}(z)W_{4}(w)\rangle ={2^{10}(k+1)(k+2)(k+3)(2k-1)(3k-4)\over
k^{4}(16k-17)}{1\over (z-w)^{8}}\;,\eqno(C.3)$$
$$\langle W_{5}(z)W_{5}(w)\rangle ={9\cdot 2^{15}\over 5}{(k+1)(k+2)(k+3)(k+4)(
2k-1)(5k-8)\over k^{5}(64k-107)}{1\over (z-w)^{10}}\;.\eqno(C.4)$$

\end{document}